\documentclass[sigconf]{acmart}

\usepackage{booktabs} 

\usepackage{amsmath,url}
\usepackage{graphicx,subcaption}
\usepackage{color}
\usepackage{CJKutf8}
\usepackage{array}
\usepackage{arydshln}
\usepackage{adjustbox}

\newif\ifworkinprogress
 \workinprogressfalse

\ifworkinprogress
	\newcommand{\ms}[1]{\textcolor{blue}{\textbf{[Markus] #1}}}
	\newcommand{\chb}[1]{\textcolor{magenta}{\textbf{[Christine] #1}}}
\else
  \newcommand{\ms}[1]{}
  \newcommand{\chb}[1]{}
\fi

\setcopyright{rightsretained}

\setcopyright{none}
\makeatletter
\renewcommand\footnotetextcopyrightpermission[1]{
    \begin{figure}[b!]
        \begin{flushleft}
            \noindent
            cite as:
            \newline
            \textit{Markus Schedl \& Christine Bauer (2017). Online Music Listening Culture of Kids and Adolescents: Listening Analysis and Music Recommendation Tailored to the Young. 1st International Workshop on Children and Recommender Systems (KidRec 2017), co-located with 11th ACM Conference on Recommender Systems (RecSys 2017), Como, Italy, 27 August.}
            \newline
            Copyright \textcopyright 2017 for this paper by its authors. Use permitted under Creative Commons License Attribution 4.0 International (CC BY 4.0).\\ \includegraphics[scale=0.8]{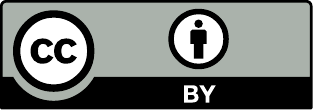}
    \end{flushleft}
    \end{figure}
}

\makeatother

\usepackage{fancyhdr}

\begin{document}
\title{Online Music Listening Culture of Kids and Adolescents} 
\subtitle{Listening Analysis and Music Recommendation Tailored to the Young}

\author{Markus Schedl}
\orcid{0000-0003-1706-3406}
\affiliation{%
  \institution{Johannes Kepler University Linz}
  \streetaddress{Department of Computational Perception}
  \city{Linz} 
  \state{Austria} 
}
\email{markus.schedl@jku.at}

\author{Christine Bauer}
\orcid{0000-0001-5724-1137}
\affiliation{%
  \institution{Johannes Kepler University Linz}
  \streetaddress{Department of Computational Perception}
  \city{Linz} 
  \state{Austria} 
}
\email{christine.bauer@jku.at}

\renewcommand{\shortauthors}{Schedl and Bauer}

\begin{abstract}
In this paper, we analyze a large dataset of user-generated music listening events from Last.fm, focusing on users aged 6 to 18 years.
Our contribution is two-fold.
First, we study the music genre preferences of this young user group and analyze these preferences for homogeneity within more fine-grained age groups and with respect to gender and countries. 
Second, we investigate the performance of a collaborative filtering recommender when tailoring music recommendations to different age groups. We find that doing so improves performance for all user groups up to 18 years, but decreases performance for adult users aged 19 years and older.
\end{abstract}

%
%


\maketitle

\fancyhead{}
\pagestyle{fancy}
\fancyhf{}
\lhead{1st International Workshop on Children \& Recommender Systems (KidRec '17), August 27, 2017, Como, Italy}
\rhead{Schedl and Bauer}
\rfoot{\thepage}

%
%

\renewcommand{\baselinestretch}{0.99}

\section{Introduction and Context}\label{sec:introduction}
Kids and adolescents use computers mainly for learning and entertainment purposes~\cite{Chiasson:2005:TME:1054972.1055089} and this age group is also the one that is highly engaged with music~\cite{ifpi2016consumerinsight}. Therefore, it does not come as a surprise that especially in the age group of $\mathrm{<}$25 years, music streaming portals are growing in popularity~\cite{ifpi2016consumerinsight}. Music service providers offering integrated music recommender systems thus have to be prepared for this young user group.

Considering user properties, including demographics such as gender, age, or country (e.g., 
\cite{al2016user,zhao2014we}), is a widely adopted approach for recommender systems and has been focus of research in the past few years.
In the field of music recommender systems, relying on listening histories or ratings is nevertheless still the most common approach~\cite{schedl_etal:fntir:2014}. 
Still, recent work (e.g.,~\cite{shi_etal:compsurv:2014,schedl_hauger:sigir:2015}) shows that integrating different listener or listening information can substantially improve the quality of music recommendations.

Studies investigating the relationship between age and music preferences are particularly rare. Most researchers draw their samples from the population of university students; hence, samples are mostly homogeneous with respect to age~\cite{laplante2014}. The few studies that allow to draw conclusions with respect to age, though, found that it is substantially associated with music preferences, particularly in terms of genre~\cite{ter2011intergenerational,harrison2010musical} and suggest to consider the relationship between age and music taste in music recommender systems~\cite{laplante2014}.

Against this background, the contribution of this paper is two-fold. First, we analyze music preferences of kids and adolescents based on the LFM-1b dataset~\cite{schedl:icmr:2016}, which aggregates information about more than one billion listening events by more than $120,\!000$ Last.fm users. 
Second, exploiting the same dataset, we study the performance of a collaborative filtering approach, tailoring its recommendations to particular age groups, ranging from 6 to 18 years.

To allow for a clear structure, we first describe the methods and material used for our studies in Section~\ref{sec:method}. In Section~\ref{sec:taste}, we present our findings on the music taste of kids and adolescents, detailing differences with respect to gender, country, and fine-grained age groups. In Section~\ref{sec:taste}, we report and discuss our findings on the recommendation experiments. Finally, we conclude with a summary and outlook to future work in Section~\ref{sec:conclusions}.


%
%

\section{Methods and Material}\label{sec:method}
In the following, we describe the dataset, our approach to model music preferences on the user level, and how we investigate a user group's \emph{preferences} and \emph{homogeneity} of these preferences.

\subsection{Dataset}\label{dataset}
\begin{figure}[t!]
\centering
\includegraphics[width=1.0\linewidth, trim=2.6cm 1.5cm 1.1cm 1.0cm]{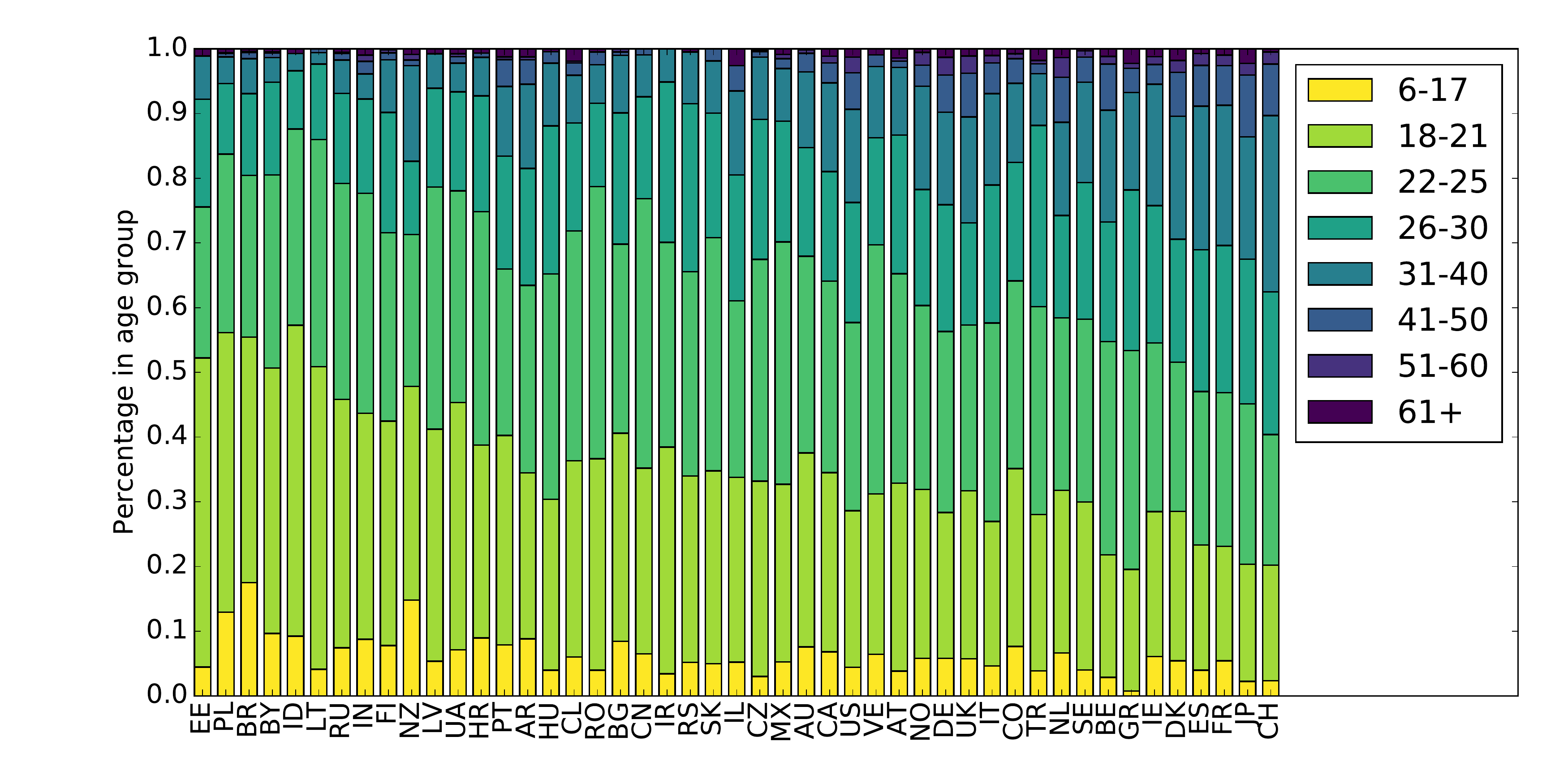}
\caption{\label{fig:age_distribution}Distribution of users over age groups, sorted according to median age from young to old, for the top countries with at least 100 users in the LFM-1b dataset.}
\end{figure}

For our analysis, we exploit the LFM-1b dataset~\cite{schedl:icmr:2016} of 1,088,161,692 individual listening events created by 120,175 users of the music platform Last.fm, who listened to 585,095 unique artists after data cleansing as described in~\cite{schedl:icmr:2016}. Out of these 120,175 users, 46,120 (38.4\%) provide age information in their profile. 
Considering only those who provide their age, users from 6 to 18 years (inclusive) represent 5,953 (12.9\%). Including users up to age 25, this number increases to almost two thirds of the population (30,404 users or 65.9\%).
Figure~\ref{fig:age_distribution} shows the distribution of age groups for the top countries in the dataset, i.e.,~those with at least 100 users.\footnote{The country abbreviations comply with the ISO 3166 standard: \url{https://www.iso.org/iso-3166-country-codes.html}}
The stacked bars are sorted according to median age from young to old (left to right). For instance, more than half of the users in Estonia, Poland, Brazil, Belarus, India, and Lithuania are younger than 22.

\subsection{Modeling and Analyzing Music Preferences and Homogeneity}
\label{modeling}
To model music preferences on a user level, we 
gather the top user-generated tags for each artist in the LFM-1b dataset, using the Last.fm API endpoint \texttt{artist.getTopTags}. 
We index the tags using a dictionary of $20$ main genres from Allmusic, 
casefold tags and index terms, and describe each artist by a bag-of-words representation of genres.
Considering each user's playcount vector over artists, we compute his or her \emph{genre profile}. To this end, each artist's genre occurrence is multiplied with the respective playcount value of the user for that artist. Summing up these playcount-weighted artists' genre occurrences on the genre level for each user results in a 20-dimensional feature vector over the 20 genres. We 
normalize these vectors for each user, so that the user's genre profile contains the percentage of music listened to from each of the 20 genres.
Based on the genre profiles, we measure music preferences for a given user group (e.g., users aged 6 to 12 years) by computing the arithmetic mean 
over all group members' genre profiles.
We further quantify the homogeneity 
of preferences within a given user group using Krippendorff's~$\alpha$ score of inter-rater agreement~\cite{krippendorff:content:2013}.

\subsection{Recommender Systems Evaluation}\label{sec:approach_recsys}



To investigate whether music recommender systems perform better when tailoring recommendations to particular age groups 
we conduct rating or preference prediction experiments, which is a common evaluation approach in recommender systems research.
We analyze the performance of a model-based collaborative filtering approach tailoring the recommendations to age groups from 6 and 18 years, and compare results with those realized for adults (aged 19 to 60 years) and the overall population.
To this end, we first normalize and scale the playcount values in the user-artist-matrix of the LFM-1b dataset to the range [0, 1000] for each user individually, assuming that higher numbers of playcounts indicate higher user preference for an artist (for the relation between implicit and explicit feedback see, e.g., \cite{parra2011,jawaheer2010comparison}).
We apply singular value decomposition according to~\cite{salakhutdinov2007probabilistic}, equivalent to probabilistic matrix factorization, to factorize the user-artist-matrix and in turn effect rating prediction. In 5-fold cross-validation experiments with random shuffle across all users, we use root mean square error (RMSE) and mean absolute error (MAE) as performance measures.

\section{Music Preferences of the Young}\label{sec:taste}
In this section, we discuss the overall preferences of young listeners (Section~\ref{young}). Then, we further detail these preferences by considering gender (Section~\ref{gender}) and country (Section~\ref{country}) information. Finally, we delve into details on music preferences of various age groups within the young listener population (Section~\ref{agegroups}).

\subsection{Overall Music Preferences}\label{young}
Table~\ref{tbl:gp} shows the arithmetic means and standard deviations (in parentheses) of the genre profiles for the entire Last.fm population (first row), for all young listeners until 18 years (second row), for all adult listeners aged 19 and older (third row), and for categories of different user groups (e.g., all user groups distinguished according to their country or according to their age). \textcolor{blue}{Blue} and \textcolor{red}{red} font is used to indicate, respectively, \textcolor{blue}{highest} and \textcolor{red}{lowest} value per genre within each category of user groups. For instance, when categorizing our target group of young listeners (aged 0 to 18) with respect to country, metal is listened to least in the US (3.20\%) and most in Poland (9.12\%) and Finland (8.87\%).
The last column of the table contains Krippendorff's agreement score~$\alpha$, which quantifies homogeneity.
Please note that we only show results for genres with an overall share among all users' listening events of at least 3\%. Detailed results for all genres can be provided by the authors upon request.

The first row of the table contains the overall genre distribution of the entire population (irrespective of age). It reveals that the top genres listened to by the entire LFM-1b sample are rock ($18.27\%$), alternative ($16.75\%$), and pop ($13.64\%$).
The second row, aggregating young listeners (up to 18 years, inclusive), 
shows that the top genres are the same as for the overall population, though the preferences for rock ($20.54\%$) and alternative ($19.03\%$) are even more pronounced than in the overall population; the opposite for pop ($12.99\%$). Furthermore, much higher preferences among the young are observed for metal ($5.96\%$ vs.~$3.98\%$) and punk ($8.53\%$ vs.~$6.19\%$), whereas substantially lower preferences exist for rnb ($2.76\%$ vs.~$3.34\%$), jazz ($2.56\%$ vs.~$3.97\%$), and blues ($2.23\%$ vs.~$3.28\%$).

A comparison of the preferences of young listeners up to 18 years (second row) and listeners aged 19 and above (third row) shows a comparable picture: The genres that are preferred more by young listeners compared to adults are, respectively, rock ($20.17\%$ vs.~$19.49\%$), alternative ($19.03\%$ vs.~$17.85\%$), pop ($12.99\%$ vs.~$12.57\%$), metal ($5.96\%$ vs.~$5.25\%$), rap ($3.66\%$ vs.~$2.78\%$), and rnb ($2.76\%$ vs.~$2.34\%$), whereas the genres preferred more by adults than by young listeners are electronic ($11.67\%$ vs.~$11.07\%$), folk ($5.76\%$ vs.~$4.73\%$), jazz ($3.67\%$ vs.~$2.56\%$), and blues ($2.89\%$ vs.~$2.23\%$).

With an overall agreement score of $\alpha=0.493$, moderate homogeneity in genre preferences can be observed for the entire user population, according to~\cite{landis_koch:1977}. Compared to most analyzed other user groups~-- with respect to age and/or country~-- this overall agreement for genre preference is rather low.

Generally, our data suggests that rock is the most preferred genre across all considered user groups except for young listeners in the United Kingdom, who slightly prefer alternative ($19.68\%$) to rock ($19.08\%$). Taking this general perspective, alternative ranks second across all user groups except for young listeners in the United Kingdom. Blues appears to be the least preferred genre, which holds true for most considered user groups except for young listeners from Poland, who like rnb less ($2.41\%$ vs. $2.22\%$), from the Netherlands who like jazz less ($2.77\%$ vs. $2.76\%$), and from Brazil who appreciate rap less than blues ($2.72\%$ vs. $2.16\%$). Further, compared to the overall population and young listeners, adults aged 19 years and older like rnb less than blues ($2.89\%$ vs. $2.34\%$).

\subsection{Gender-specific Music Preferences}\label{gender}
According to our data, rock is the genre most listened to by both male ($20.47\%$) and female ($19.49\%$) young listeners. Blues ($2.17\%$), rnb ($2.49\%$) and jazz ($2.55\%$) are the least preferred genres for the male users in this age group; blues ($2.29\%$), jazz ($2.50\%$) and rap ($2.94\%$) for females.
Further, our data suggests a substantial male preference for metal ($7.22\%$ vs.~$4.40\%$) and rap ($4.21\%$ vs.~$2.94\%$); pop ($14.71\%$ vs.~$11.71\%$) is particularly preferred by female users.

The homogeneity of music taste with respect to genre is substantially higher for females ($\alpha=0.630$) than for males ($\alpha=0.496$).
In fact, the homogeneity is higher for the female user group than for any other user group considered in our analysis (Table~\ref{tbl:gp}).

\subsection{Country-specific Music Preferences}\label{country}
The general preference for rock music among young listeners seems to be consistent across all analyzed countries. A similar picture is shown for the genres alternative and pop. 
Country-specific differences can be seen for other genres, though. 

For instance, in Poland ($9.12\%$) and Finland ($8.87\%$) the liking of metal is particularly high compared to other countries score for metal. Metal is also the genre that shows the highest gap in preference between countries, with Polish listeners being most affine ($9.12\%$) and US listeners liking this genre least ($3.20\%$). 
Other substantial discrepancies between countries are observed for pop with highest share in Sweden ($15.90\%$) and lowest in Russia ($10.96\%$), for electronic with highest share in Russia ($14.26\%$) and lowest in Brazil ($8.31\%$), for alternative with highest share in Poland ($19.83\%$) and lowest in Finland ($16.56\%$), for rnb with highest share in the United Kingdom ($3.51\%$) and lowest in Russia ($1.82\%$), and for rap with highest share in Germany ($5.60\%$) and lowest in Brazil ($2.16\%$).

The highest homogeneity of music preferences can be found for the United Kingdom ($\alpha=0.623$) and Sweden ($\alpha=0.612$), which are higher compared to the overall group of young users ($\alpha=0.539$) and the overall user population ($\alpha=0.493$).

\subsection{Music Preferences in Different Age Groups}
\label{agegroups}
Comparing the genre preferences of different age groups within the young listener population, our data suggests that the young listener's high preference for rock music and the rather low preference for blues holds also for the more fine-grained user groups.


Our data further suggests that rnb ($3.66\%$), rap ($4.35\%$), blues ($3.46\%$), and jazz ($4.02\%$) are most liked by the youngest age group (6,12), although overall with rather low listening shares compared to other genres. The youngest age group (6,12) also appreciates electronic music ($12.92\%$) the most in comparison to the other age groups, in this case with considerable preference scores.
In contrast, rock ($18.85\%$), folk ($4.41\%$), punk ($6.60\%$), alternative ($17.18\%$), and metal ($4.06\%$) are least liked by the youngest group, compared to the older groups. A preference for these genres evolves, however, with increasing age up to 16 years; then it steadily decreases.

Furthermore, results indicate that the preference for folk music tends to rise with increasing age (from $4.41\%$ to $4.81\%$).
The liking of rnb ($4.24\%$), rap ($4.47\%$), and pop ($13.49\%$) reach their peak scores for the age group (13,14). Preference for rock ($20.44\%$), punk ($9.01\%$), alternative music ($19.26\%$), and metal ($6.11\%$) peaks for the age group (15,16); then, the preference scores decrease with increasing age.
The opposite is observed for other genres: Preference for electronic music ($10.60\%$) and jazz ($2.23\%$) scores lowest for the age group (15,16), for blues ($2.02\%$) for the age group (13,14); the preference for these genres tends to rise with increasing age.




\begin{table}[t!p]
\begin{adjustbox}{angle=90}
\centering\tabcolsep=0.075cm
\scalebox{.95}{
\begin{tabular}{|c|c|c||r||r|r|r|r|r|r|r|r|r|r|r|r|} 
\hline
cty & age & gender & no.~users & rnb & rap & elect. & rock & blues & folk & jazz & punk & altern. & pop & metal & $\alpha$ \\
\hline
- & - & - & 120175 & 3.34 (1.42) &  3.41 (1.44) &  11.18 (0.74) &  18.27 (0.35) &  3.28 (1.01) &  5.61 (0.75) &  3.97 (0.85) &  6.19 (0.89) &  16.75 (0.33) &  13.64 (0.43) &  3.98 (1.76) &  0.493 \\
\hline
- & (0,18) & - & 6116 &  \bf{\textcolor{blue}{2.76}} (4.31)  &  \bf{\textcolor{blue}{3.66}} (5.48)  &  \bf{\textcolor{red}{11.07}} (7.60)  &  \bf{\textcolor{blue}{20.17}} (6.45)  &  \bf{\textcolor{red}{2.23}} (2.39)  &  \bf{\textcolor{red}{4.73}} (4.13)  &  \bf{\textcolor{red}{2.56}} (2.52)  &  \bf{\textcolor{blue}{8.53}} (6.42)  &  \bf{\textcolor{blue}{19.03}} (5.26)  &  \bf{\textcolor{blue}{12.99}} (6.30)  &  \bf{\textcolor{blue}{5.96}} (8.82)  &  0.539  \\
- & (19,60) & - & 39553 &  \bf{\textcolor{red}{2.34}} (3.58)  &  \bf{\textcolor{red}{2.78}} (4.08)  &  \bf{\textcolor{blue}{11.67}} (7.90)  &  \bf{\textcolor{red}{19.49}} (6.04)  &  \bf{\textcolor{blue}{2.89}} (2.81)  &  \bf{\textcolor{blue}{5.67}} (4.17)  &  \bf{\textcolor{blue}{3.67}} (3.00)  &  \bf{\textcolor{red}{7.06}} (5.68)  &  \bf{\textcolor{red}{17.85}} (4.91)  &  \bf{\textcolor{red}{12.57}} (5.64)  &  \bf{\textcolor{red}{5.25}} (8.11)  &  0.546  \\
\hline  

- & (0,18) & m & 3324 &  \bf{\textcolor{red}{2.49}} (4.19)  &  \bf{\textcolor{blue}{4.21}} (6.33)  &  \bf{\textcolor{blue}{11.41}} (8.55)  &  \bf{\textcolor{blue}{20.54}} (6.93)  &  \bf{\textcolor{red}{2.17}} (2.52)  &  \bf{\textcolor{red}{4.32}} (4.14)  &  \bf{\textcolor{blue}{2.55}} (2.60)  &  \bf{\textcolor{blue}{8.75}} (6.88)  &  \bf{\textcolor{red}{18.67}} (5.43)  &  \bf{\textcolor{red}{11.71}} (6.32)  &  \bf{\textcolor{blue}{7.22}} (9.85)  &  0.496  \\
- & (0,18) & f & 2362 &  \bf{\textcolor{blue}{3.14}} (4.47)  &  \bf{\textcolor{red}{2.94}} (3.99)  &  \bf{\textcolor{red}{10.42}} (5.79)  &  \bf{\textcolor{red}{19.77}} (5.70)  &  \bf{\textcolor{blue}{2.29}} (2.18)  &  \bf{\textcolor{blue}{5.20}} (3.94)  &  \bf{\textcolor{red}{2.50}} (2.36)  &  \bf{\textcolor{red}{8.40}} (5.80)  &  \bf{\textcolor{blue}{19.66}} (4.93)  &  \bf{\textcolor{blue}{14.71}} (5.73)  &  \bf{\textcolor{red}{4.40}} (7.04)  &  0.630  \\
\hline 

BR & (0,18) & - & 901 &  3.25 (4.98)  &  \bf{\textcolor{red}{2.16}} (2.76)  &  \bf{\textcolor{red}{8.31}} (5.86)  &  19.98 (6.00)  &  2.72 (2.43)  &  \bf{\textcolor{blue}{5.49}} (4.05)  &  2.63 (2.57)  &  8.79 (6.02)  &  19.82 (4.86)  &  14.42 (6.26)  &  5.95 (8.52)  &  0.595  \\
PL & (0,18) & - & 801 &  2.22 (3.58)  &  4.60 (7.85)  &  9.10 (7.14)  &  20.59 (7.33)  &  2.41 (2.54)  &  4.16 (3.54)  &  2.54 (2.55)  &  8.69 (6.53)  &  \bf{\textcolor{blue}{19.83}} (5.50)  &  10.97 (6.51)  &  \bf{\textcolor{blue}{9.12}} (10.77)  &  0.495  \\
US & (0,18) & - & 640 &  3.01 (4.55)  &  3.96 (5.00)  &  11.33 (6.58)  &  19.60 (5.57)  &  2.09 (2.27)  &  5.25 (3.79)  &  2.72 (2.46)  &  9.44 (7.00)  &  19.01 (4.96)  &  14.48 (5.77)  &  \bf{\textcolor{red}{3.20}} (6.22)  &  0.602  \\
RU & (0,18) & - & 592 &  \bf{\textcolor{red}{1.82}} (3.14)  &  3.94 (5.72)  &  \bf{\textcolor{blue}{14.26}} (8.60)  &  21.33 (6.46)  &  1.79 (2.15)  &  3.70 (4.03)  &  2.03 (1.88)  &  8.95 (6.26)  &  19.43 (5.17)  &  \bf{\textcolor{red}{10.96}} (5.50)  &  5.95 (7.42)  &  0.601  \\
DE & (0,18) & - & 340 &  2.68 (4.20)  &  \bf{\textcolor{blue}{5.60}} (7.07)  &  12.36 (7.84)  &  19.95 (6.76)  &  \bf{\textcolor{red}{1.36}} (1.64)  &  4.56 (5.24)  &  \bf{\textcolor{red}{1.82}} (2.01)  &  \bf{\textcolor{blue}{9.47}} (6.74)  &  17.77 (5.61)  &  11.99 (6.26)  &  6.48 (9.62)  &  0.489  \\
UK & (0,18) & - & 337 &  \bf{\textcolor{blue}{3.51}} (4.90)  &  3.19 (4.19)  &  12.47 (7.91)  &  \bf{\textcolor{red}{19.08}} (5.26)  &  2.01 (2.36)  &  4.35 (3.19)  &  2.54 (2.55)  &  8.53 (5.67)  &  19.68 (4.86)  &  15.00 (5.50)  &  3.56 (6.38)  &  0.623  \\
FI & (0,18) & - & 155 &  2.56 (3.60)  &  4.85 (6.09)  &  12.19 (8.23)  &  \bf{\textcolor{blue}{21.95}} (7.02)  &  1.51 (2.02)  &  \bf{\textcolor{red}{3.46}} (3.28)  &  1.96 (2.16)  &  8.03 (7.06)  &  \bf{\textcolor{red}{16.56}} (5.54)  &  12.23 (6.77)  &  8.87 (10.09)  &  0.511  \\
NL & (0,18) & - & 135 &  2.87 (4.56)  &  3.70 (5.74)  &  11.07 (8.56)  &  19.61 (7.18)  &  \bf{\textcolor{blue}{2.77}} (2.72)  &  5.13 (4.70)  &  2.76 (2.33)  &  7.95 (7.84)  &  18.27 (5.60)  &  13.35 (5.68)  &  5.61 (9.16)  &  0.476  \\
SE & (0,18) & - & 90 &  2.36 (2.92)  &  2.44 (2.32)  &  12.45 (6.90)  &  19.65 (5.45)  &  2.07 (2.23)  &  5.20 (3.56)  &  \bf{\textcolor{blue}{3.02}} (2.95)  &  \bf{\textcolor{red}{7.38}} (6.21)  &  18.46 (5.14)  &  \bf{\textcolor{blue}{15.90}} (8.10)  &  4.08 (6.73)  &  0.612  \\
\hline

- & (6,12) & - & 81 &  3.66 (5.02)  &  4.35 (6.52)  &  \bf{\textcolor{blue}{12.92}} (9.24)  &  \bf{\textcolor{red}{18.85}} (7.22)  &  \bf{\textcolor{blue}{2.46}} (2.26)  &  \bf{\textcolor{red}{4.41}} (2.91)  &  \bf{\textcolor{blue}{4.02}} (3.08)  &  \bf{\textcolor{red}{6.60}} (5.30)  &  \bf{\textcolor{red}{17.18}} (4.92)  &  13.10 (6.07)  &  \bf{\textcolor{red}{4.06}} (6.93)  &  0.490  \\
- & (13,14) & - & 258 &  \bf{\textcolor{blue}{4.24}} (5.60)  &  \bf{\textcolor{blue}{4.47}} (6.49)  &  11.01 (7.64)  &  19.06 (6.36)  &  \bf{\textcolor{red}{2.02}} (2.22)  &  4.46 (3.94)  &  2.48 (2.78)  &  7.77 (6.43)  &  18.73 (5.53)  &  \bf{\textcolor{blue}{13.49}} (6.36)  &  5.62 (9.16)  &  0.491  \\
- & (15,16) & - & 1439 &  2.91 (4.41)  &  \bf{\textcolor{red}{3.58}} (5.26)  &  \bf{\textcolor{red}{10.60}} (7.27)  &  \bf{\textcolor{blue}{20.44}} (6.47)  &  2.11 (2.37)  &  4.48 (4.23)  &  \bf{\textcolor{red}{2.23}} (2.32)  &  \bf{\textcolor{blue}{9.01}} (6.64)  &  \bf{\textcolor{blue}{19.26}} (5.41)  &  13.40 (6.68)  &  \bf{\textcolor{blue}{6.11}} (8.85)  &  0.547  \\
- & (17,18) & - & 4190 &  \bf{\textcolor{red}{2.60}} (4.14)  &  3.63 (5.48)  &  11.14 (7.62)  &  20.22 (6.40)  &  2.25 (2.38)  &  \bf{\textcolor{blue}{4.81}} (4.09)  &  2.58 (2.47)  &  8.51 (6.35)  &  19.06 (5.17)  &  \bf{\textcolor{red}{12.85}} (6.16)  &  6.03 (8.87)  &  0.544  \\
\hline

\end{tabular}
}
}
\caption{\label{tbl:gp}Music preferences for age groups.} 

\end{adjustbox}

\end{table}

\section{Music Recommendation Experiments}\label{sec:recsys}
We conduct preference prediction experiments for various age groups as described in Section~\ref{sec:approach_recsys} and report error measures in Table~\ref{tab:results}.
An overall performance score is obtained using all user playcounts of the dataset, independent of the users' age (first row). To assess to which extent tailoring recommendations to different age groups affects recommendation performance, we create subsets of users according to their membership in age groups 6-12, 13-14, 15-16, and 17-18; then we perform the same experiment as described above individually on these subsets. The results for a subset comprising the entire group of 6- to 18-aged users can be found in the second row of Table~\ref{tab:results}. The third row contrasts these results to the user group of adults (19 to 60 years). The results for the more fine-grained age ranges can be found in the bottom rows.
Our discussion focuses on the RMSE values; 
the insights gained from the RMSE values correspond to the ones 
gained via MAE. 

Our results suggest that the general performance for the whole young group (0,18) substantially differs from that of the overall population (RMSE of $7.766$ vs.~$29.105$). 
In addition, RMSE is  smaller for all age groups $\mathrm{\le}$18 years compared to the error for the overall population. This indicates that kids and adolescents aged 6 to 18 benefit substantially from like-minded peers when recommending items with collaborative filtering, as underpinned by RMSE values as low as $5.178$ to $10.395$. This observation is in line with findings from development psychology that music is considered a means for socializing with peers during adolescence~\cite{laiho2004psychological}.
The recommendations work particularly well for the youngest age group (6,12) with an RMSE of $5.178$ and for users late in their adolescence (17,18) with an RMSE of $7.469$.

\begin{table}[t!]
\centering
\begin{tabular}{|l|r|r|r|}
\hline
\textbf{groups} & \textbf{no.~users} & \textbf{RMSE} & \textbf{MAE} \\
\hline
All users & 120157 & 29.105 & 25.202 \\
\hline
All young users (0,18) & 6101				& 7.766	& 2.940 \\
All adult users (19,60) & 39514 			& 77.548 & 76.131 \\
\hdashline[0.5pt/1pt]
(6,12) 	& 80 	& 5.178 	& 1.555 \\
(13,14)	& 257 	& 10.395	& 4.230 \\
(15,16) & 1435	& 9.513		& 3.815 \\
(17,18)	& 4181	& 7.469		& 2.835 \\


\hline
\end{tabular}
\caption{Error measures for different age groups, with playcounts scaled to [0, 1000].}
\label{tab:results}
\end{table}

\section{Conclusions and Future Work}\label{sec:conclusions}
We analyzed the music preferences of kids and adolescents aged 6 to 18 years in terms of genre preferences and homogeneity of these preferences, based on the LFM-1b dataset of Last.fm users. We uncovered substantial differences in both preferences and homogeneity between young users, adult users, and the overall user population. Such differences were also found between countries and gender of the young population and between fine-grained age groups.
In recommender systems experiments, 
we found that preference predictions were substantially more accurate for the young user groups than for the adult population. We conclude that tailoring a collaborative filtering systems to users $\mathrm{\le}$18 years is beneficial.

A limitation of our approach is that the LFM-1b dataset may not necessarily generalize to the population at large, in particular in terms of age distribution.
Still, as listeners up to 18 years are well represented in the dataset and this age group is known to use social media platforms frequently~\cite{chassiakos2016children}, we assume that the dataset provides a good indicator. Further in-depth investigation is necessary, especially 
with respect to the highly varying ``music listening culture" in different countries. We will integrate more data sources 
and deploy additional research instruments (e.g., surveys).

\section*{Acknowledgments}
This research is supported by the Austrian Science Fund (FWF): V579 and P25655.

\bibliographystyle{ACM-Reference-Format}
\bibliography{kidrec2017}

\end{document}